\documentclass[sigconf]{acmart}

\usepackage{booktabs} 

\setcopyright{acmcopyright}

\usepackage{amsmath}

\usepackage{natbib}
\usepackage{graphicx,float}
\usepackage{textcomp}
\usepackage{xcolor}
\usepackage{csquotes}
\usepackage{subcaption}
\usepackage{lipsum}
\usepackage{tabularx,multirow,booktabs,blindtext}

\usepackage{colortbl}
\usepackage{xcolor}

\usepackage{algorithm}
\usepackage{algpseudocode}
\algblock{Input}{EndInput}
\algnotext{EndInput}
\algblock{Output}{EndOutput}
\algnotext{EndOutput}

\usepackage{float} 
\usepackage{tabularx}
\usepackage{ragged2e} 
\usepackage{xcolor}

\definecolor{lowcolor}{RGB}{255, 0, 0}

\definecolor{mediumcolor}{RGB}{255,165,0}

\definecolor{highcolor}{RGB}{0, 128, 0}

\usepackage{soul,color}

\copyrightyear{2024}
\acmYear{2024}
\setcopyright{rightsretained}
\acmConference[SAC '24]{The 39th ACM/SIGAPP Symposium on Applied Computing}{April 8--12, 2024}{Avila, Spain}
\acmBooktitle{The 39th ACM/SIGAPP Symposium on Applied Computing (SAC '24), April 8--12, 2024, Avila, Spain}\acmDOI{10.1145/3605098.3636164}
\acmISBN{979-8-4007-0243-3/24/04}

\begin{document}
\title{The Performance of Sequential Deep Learning Models in Detecting Phishing Websites Using Contextual Features of URLs}

  
\renewcommand{\shorttitle}{The Performance of Sequential Deep Learning Models in Detecting Phishing Websites Using Contextual Features of URLs}

\author{Saroj Gopali, Akbar S. Namin}
\affiliation{%
  \institution{Department of Computer Science}
  \institution{Texas Tech University}
 \country{USA}
}
\email{{saroj.gopali, akbar.namin}@ttu.edu}

\author{Faranak Abri}
\affiliation{%
  \institution{Department of Computer Science}
  \institution{San Jose State University}
 \country{USA}
}
\email{faranak.abri@sjsu.edu}

\author{Keith S. Jones}
\affiliation{%
  \institution{Department of Psychology}
  \institution{Texas Tech University}
 \country{USA}
}
\email{keith.s.jones@ttu.edu}


\renewcommand{\shortauthors}{G. Saroj et al.}

\begin{abstract} 
Cyber attacks continue to pose significant threats to individuals and organizations, stealing sensitive data such as personally identifiable information, financial information, and login credentials.  Hence, detecting malicious websites before they cause any harm is critical to preventing fraud and monetary loss. To address the increasing number of phishing attacks, protective mechanisms must be highly responsive, adaptive, and scalable. Fortunately, advances in the field of machine learning, coupled with access to vast amounts of data, have led to the adoption of various deep learning models for timely detection of these cyber crimes.  This study focuses on the detection of phishing websites using deep learning models such as Multi-Head Attention, Temporal Convolutional Network (TCN), BI-LSTM, and LSTM where URLs of the phishing websites are treated as a sequence. The results demonstrate that Multi-Head Attention and BI-LSTM model outperform some other deep learning-based algorithms such as TCN and LSTM in producing better precision, recall, and F1-scores.

\end{abstract}
%
%



\maketitle

\vspace{-0.2cm}
\section{Introduction}

Phishing is a type of cyber attacks in which attackers use deceptive tactics such as fake websites or emails to trick people into disclosing sensitive information such as usernames, passwords, and financial information. Phishing attacks are becoming more sophisticated and attackers are constantly creating new ways to make their scams appear more legitimate.
According to a report published by the Federal Bureau of Investigation (FBI) in 2021 \cite{ic3}, social engineering attacks such as phishing, vishing, smishing, and pharming collectively targeted $323,972$ reported victims. The social engineering attacks grew by approximately 34.5\% from 2020 to 2021, making it the crime with the most victims. In 2022, there was a decrease of approximately $7.3\%$ from the previous year's 323,972 victims to 300,497 in social engineering attacks. However, even with the decrease, social engineering attacks continue to have a relatively high number of victims compared to other reported crimes that resulted in a total loss of \$52 million US Dollars \cite{ic3}. 

With the advances in Artificial Intelligence (AI), the application of machine learning and deep learning in cybersecurity has gained significant attention and has been extensively used to tackle security-related problems.
Jain and Gupta \cite{jain2016novel} proposed white-list-based methods for suspicious URLs to protect against phishing attacks. Chatterjee and Namin \cite{8754075} proposed a reinforcement learning-based framework for detecting phishing websites. Yang et al.\cite{yang2019phishing} proposed an approach for detecting phishing URLs a multidimensional feature method that utilizes fast CNN-LSTM deep learning techniques (MFPD) to accurately identify phishing URLs promptly. Recently, Otieno et al.\  \cite{Denish1, Denish2} introduced a transformer-based for detecting phishing emails. 

Unlike existing research, which relied mainly on static features of Web pages and URL analysis, this paper present a novel comparative analysis of end-to-end deep learning algorithms for detecting phishing websites straight from a given URL link. The novelty lies in our approach in URLs treated as a ``{\it sequence}'' of tokens enabling adaptation of algorithms developed for sequence analysis. We build and test neural networks such as Temporal Convolutional Networks (TCN), Long Short-Term Memory (LSTM), and its bidirectional variation (BiLSTM) as well as Multi-Head Attention-based networks. To the best of our knowledge, this is the first time these cutting-edge deep learning algorithms have been utilized to analyze URL sequences and their representations for universal phishing Websites detection.

Our findings indicate that all four deep learning models perform similarly and surpass traditional feature-based phishing detection methods that rely on URL syntactical features (i.e., not sequential features). 
Notably, the Multi-Head Attention and BiLSTM models outperform the other two models. These results underscore the effectiveness of deep learning techniques in identifying phishing Websites and emphasize their potential in practical applications. Additionally, we report the training time and duration for these deep learning-based methods, offering a cost-effective analysis of their implementation and deployment.

\vspace{-0.2cm}
\section{URL As A Sequence}
\label{sec:URLsequence}

In the past, analyzing the context of data relied heavily on engineering syntactic features \cite{fahmi2006learning}. But today contextual information has transitioned to analyzing sequential data \cite{nadkarni2011natural}. This shift is especially evident in URL analysis, where techniques such as n-grams analysis and some other sophisticated models like Recurrent Neural Networks (RNN) \cite{bahnsen2017classifying} and transformers translate the problem into a decoding problem \cite{maslej2020comparison}. In this paper, we view a given URL  as a sequence of tokens, which allows us to adapt the sequence analysis techniques for the problem at hand. 

\vspace{-0.2cm}
\section{Deep Learning Models}
\label{sec:deeplearningmodels}


\begin{algorithm}[t]
    \caption{URL-based Phishing Website Detection using Deep Learning Algorithms.}
        \label{alg:phishing-detection}
    \begin{algorithmic}[1]
        \Require{Dataset of URLs labeled as phishing or legitimate}
        \Ensure{Trained deep learning model for phishing website detection}
        \Function{TrainModel}{$URLs$}
            \State Convert URLs data into numerical vectors using embedding 
            \State Split data into training (80$\%$) and testing (20$\%$) sets
            \State Build deep learning model architecture (i.e. Multi-Head Attention, TCN, BI-LSTM, LSTM)
            \State Train model on the training set
            \State Evaluate model performance on the testing set
        \EndFunction
        
        \Function{Predict}{$model, testing\_url$}            
            \State Predict the probability of $ testing\_url$ being a phishing website using a trained deep-learning model
            \State \textbf{Return} 1 if predicted probability $>= 0.5$ else return 0.
        \EndFunction
        \Function{Performance Matrix}{true value, predicted value}
            \State Compute the confusion matrix from true value and predicted values
            \State Measure the true positive (TP), false positive (FP), true negative (TN), and false negative (FN) rates from the confusion matrix
            \State Calculate the accuracy, precision, recall, and F1 score from the TP, FP, TN, and FN rates
            \State \textbf{Return} the performance metrics
        \EndFunction
    \end{algorithmic}
\end{algorithm}

Algorithm \ref{alg:phishing-detection} lists the procedures that takes the URL address of a website. The URL is tokenized using keras' tokenization supporting libraries. The converted numerical representations by tokenizer then is fed into a deep learning model to process and predict. If the predicted probability is $\geq 0.5$, it is classified as a phishing Website, otherwise not.

During the models training, the batch size and epochs were set to 32 and 10, respectively.  These values are set simultaneously to ensure consistency across the models built and studied. The models are compiled using Adam as the optimizer and {\tt binary\_crossentropy} as the loss function. The models have different units and layer structures, which are outlined in the Table \ref{tab:dl_arch_1}. 
The models do not share any parameters and are trained separately on the same input dataset.

\begin{table}[htbp]
    \centering
   \scalebox{1}{
        \begin{tabular}{|c|l|}
        \hline
        \multicolumn{1}{|c|}{\bf Model} &  \multicolumn{1}{c|}{\bf Architecture (Layers)}  \\ 
        \hline
        Multi-Head  & \textit{1. Position Embedding} \\
         & \textit{2. Transformer Block ($Embedd\_Dimension=50, $} \\
         & \quad \textit{$num\_heads=2 ,Hidden_layer=4$)} \\
         & \textit{3. Global Average Pooling} \\
         & \textit{4. Dropout (0.04)} \\ 
        \hline
        TCN & \textit{1. Embedding } \\
         & \textit{2. TCN (unit = 126, activation =$'tanh'$)} \\
         & \textit{3. Dropout (0.04)} \\
        \hline
         LSTM & \textit{1. Embedding } \\
         & \textit{2. LSTM (unit = 256, activation = $'tanh'$)} \\
         & \textit{3. Dropout (0.04)} \\
        \hline
         BiLSTM &\textit{1. Embedding } \\
         & \textit{2. BiLSTM(unit = 35, activation =$'tanh'$)} \\
         & \textit{3. Dropout (0.04)} \\
        \hline
        \end{tabular}
    }
       \caption{Summary of deep learning architectures.}
    \label{tab:dl_arch_1}
    \vspace{-10mm}
\end{table}


The model architectures were established following a thorough exploration of hyperparameter tuning. To discover the best balance between model capabilities, we tried embedding dimensions ranging from 50 to 200, increasing by 50 at each step. Based on our experiments, we chose an embedding dimension of 50.  On the validation set, embedding sizes of 100, 150, and 200 revealed overfitting tendencies, whereas 50 demonstrated enhanced generalization.

\vspace{-0.2cm}
\section{Experimental Setup}
\label{sec:expriment}

The experiments were conducted on Google Colab and made use of GPUs in real-time. Keras library \cite{chollet2015keras} was used to build the deep learning models. For preprocessing and performance evaluation, the sklearn library was used, while matplotlib was used for plotting.

The dataset for training and testing were collected from a public Github repository\cite{ebubekirbbr}. The dataset contains a total of $73,575$ URLs, including $36,400$ legitimate URLs and $37,175$ phishing URLs. The dataset is split into a training set of $58,860$ URLs and a test set of $14,715$ URLs. The training and testing datasets contain both legitimate URLs (labeled as 0) and phishing URLs (labeled as 1). 
The URLs data tokenized using the  {\tt tf.keras.preprocessing.text} module, where the parameter {\tt num\_words} was set to $10,000$ to limit the number of unique tokens in the vocabulary. Next, the {\tt texts\_to\_sequences} method was called on the tokenizer object to convert the text into sequences of numerical tokens. 

The {\tt pad\_sequences} function was then used to pad the sequences with zeros to ensure they are all of the same lengths. The {\tt maxlen} parameter was set to $100$ to ensure that all sequences had a length of $100$. Finally, the output labels were converted to an {\tt array} format and fed to the models during training.

\vspace{-0.2cm}
\section{Results}
\label{sec:result}

The results of all models including Average Precision, Average Recall, Average F1- Score, and Average Accuracy are reported in Table \ref{table:class_report}. All models achieved a precision, recall, and F1-score average above 0.97, except for the BiLSTM model, which achieved an average score of 0.98. As per the results in Table \ref{table:class_report}, it is distinct that the TCN Model has the lowest Average ROC of 0.974 and the LSTM has the highest average ROC of 0.979, even though the distinction seems to be very minimal. Furthermore, the DQN model has demonstrated the lowest precision, recall, F1-score, and accuracy of 0.867, 0.880, 0.873, and 0.901, respectively, whereas the BiLSTM model has the highest of 0.980. The average ROC value for all models was above 0.974 except for DQN.

\subsection{Training Time}

Figure \ref{fig:training_time} displays the training time for four models (Multi-Head Attention, TCN, BiLSTM, and LSTM) at various epochs (10 to 100). The Multi-Head Attention model was trained in 5 minutes and 33 seconds at epoch 10, while the TCN model took 4 minutes and 25 seconds. The BiLSTM and LSTM models completed the training task in 4 minutes and 27 seconds and 3 minutes and 27 seconds, respectively. The Multi-Head Attention model took 31 minutes and 38 seconds to train at epoch 100; whereas, the TCN model took 33 minutes and 25 seconds. The BiSTM and LSTM models completed the task in 29 minutes and 26 seconds and 21 minutes and 24 seconds, respectively. As a result, the LSTM model requires the least amount of training time of the four models, while the TCN model requires the most.

\begin{table}[t]
    \centering
     \scalebox{0.75}{
    \begin{tabular}{|l|c|c|c|c|c|}
       
        \hline
         \multicolumn{1}{|c|}{\bf Model} & \textbf{Average} & \textbf{Average} &  \textbf{Average} &  \textbf{Average} & \textbf{Average}\\ 
        
        & \textbf{Precision} & \textbf{Recall} & \textbf{F1-score} & \textbf{Accuracy} &\textbf{ROC}\\ \hline   
        Multi-Head Attention & 0.979 & 0.979 & 0.979 & 0.980 &0.976 \\ \hline
        TCN  & 0.974 & 0.974 & 0.974 & 0.980 &0.974 \\ \hline
        LSTM & 0.977 & 0.977 & 0.977 & 0.970 &0.979 \\ \hline
        BI-LSTM & 0.980 & 0.980 & 0.980 & 0.980 &0.978 \\ \hline  
        DQN \cite{8754075} & 0.867 & 0.880 & 0.873 & 0.901 &- \\ \hline
    \end{tabular}
    }
    \caption{Average classification performance.}
    \label{table:class_report}
    \vspace{-8mm}
   
\end{table}

\begin{figure}[htbp]   
      \centering
          \vspace{-5mm}
     \begin{subfigure}[b]{\textwidth}
         \includegraphics[width=0.53\textwidth]{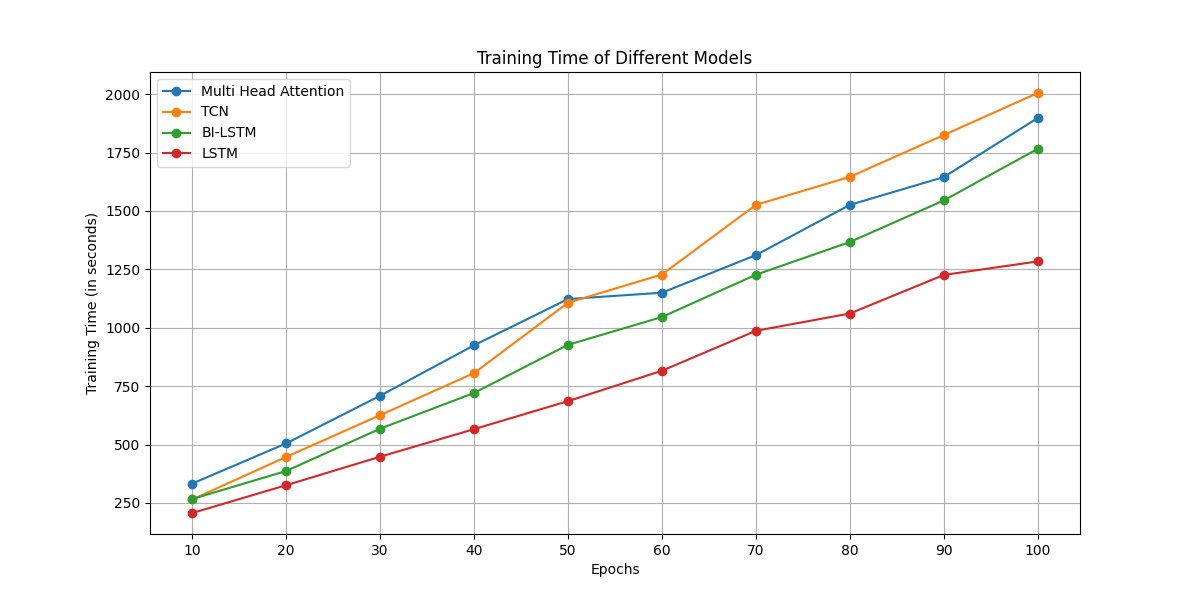}
        
         \label{fig:training_time}
     \end{subfigure}
     \centering
    \caption{ \footnotesize{Training time of different models.}}
    \vspace{-5mm}
    \label{fig:training_time}
\end{figure}

\subsection{Confusion Matrix}

As shown in Table \ref{fig:confusion} and the Confusion matrices for the LSTM, BiLSTM, TCN, and Multi-Head Attention models in Epoch 10, the LSTM model demonstrated the highest false negative value of 264, while the true negative value of 89 was the lowest across all models. The BiLSTM model, on the other hand, showed the lowest false negative value of 128 and highest True Negative of 178 among all models. The true positive value for the TCN model was 7310, and the false positive value was 7070. Similarly, the true positive and false positive values for the Multi-Head Attention model were 7,310 and 7060, respectively.

\vspace{-0.2cm}
\section{Conclusion}
\label{sec:conclusion}

This paper investigated the idea of treating URLs as sequences and leveraging sequential deep learning algorithms in detecting phishing URLs. The experiments and results analysis indicate that all four deep learning models (i.e. Multi-head Attention network, TCN, LSTM, and BiLSTM) are effective for detecting phishing websites. However, the BiLSTM model outperformed the other models, with average precision, recall, F1-score, and accuracy values of $0.980$. On the other hand, the DQN model performed the lowest, with scores of $0.867$, $0.880$, $0.873$, and $0.901$. The BiLSTM and LSTM models also outperformed the Multi-Head Attention and TCN models with an average ROC value of over $0.974$.

The TCN and LSTM Models required the most and least training time simultaneously.  Our results, show that the BiLSTM model is the most effective for end-to-end phishing Website detection directly from raw sequential URL textual data, outperforming the traditional feature-based models. The Multi-Head Attention model also shows promise as an efficient deep-learning technique for this task. 

\begin{table}[t]
    \centering
    \scalebox{0.8}{
    \begin{tabular}{|l|c|c|c|c|c|}
         \hline
        \textbf{Model}&\textbf{TN} & \textbf{FP} & \textbf{FN} & \textbf{TP} \\
   \hline
        \textbf{Multi-Head Attention Model} & 7103 & 177 & 150 & 7,285 \\
        \textbf{TCN Model} & 7060 & 220 & 125 & 7,310 \\
        \textbf{BiLSTM Model} & 7152 & 128 & 178 & 7,257 \\
        \textbf{LSTM Model} & 7016 & 264 & 89 & 7,346 \\
         \hline
    \end{tabular}
    }   
    \caption{Confusion Matrix of Models across Epochs 10: TN (True Negative), FP: (Fale Positive), FN: (False Negative), TP: (True Positive).}
    \vspace{-8mm}
    \label{fig:confusion}
\end{table}

This comparative study demonstrates the feasibility of using deep learning for generalized phishing detection in a completely featureless manner (i.e., implicit semantic-based and sequential features), providing new directions for applying these techniques to security challenges. Future research work can build on these findings by evaluating the performance of these and other deep-learning models on additional phishing website datasets.

\vspace{-0.2cm}
\section*{Acknowledgement}
This research was supported by the U.S. National Science Foundation (Awards\#: 2319802 and 2319803) and by the U.S. Office of Naval Research (Award\#: N00014-21-1-2007). Opinions, findings, and conclusions are those of the authors and do not necessarily reflect the views of the NSF or the ONR.

\vspace{-0.2cm}
\bibliographystyle{ACM-Reference-Format}
\bibliography{reference}

\end{document}